\documentclass[%
reprint,
superscriptaddress,
 amsmath,amssymb,
 aps
]{revtex4-2}
\usepackage{graphicx}% Include figure files
\usepackage{dcolumn}% Align table columns on decimal point
\usepackage{bm}% bold math
\usepackage{hyperref}% add hypertext capabilities

\usepackage{xcolor}
\usepackage{soul}

\begin{document}

\preprint{APS/123-QED}

\title{
Shape-tension coupling produces nematic order in an epithelium vertex model
}

\author{Jan Rozman}
\email{jan.rozman@physics.ox.ac.uk}
\affiliation{Rudolf Peierls Centre For Theoretical Physics, University of Oxford, Oxford OX1 3PU, United Kingdom}
\author{Rastko Sknepnek}%
\affiliation{School of Science and Engineering, University of Dundee, Dundee DD1 4HN, United Kingdom}
\affiliation{School of Life Sciences, University of Dundee, Dundee DD1 5EH, United Kingdom}
\author{Julia M. Yeomans}%
\affiliation{Rudolf Peierls Centre For Theoretical Physics, University of Oxford, Oxford OX1 3PU, United Kingdom}

\date{\today}

\begin{abstract}

We study the vertex model for epithelial tissue mechanics extended to include coupling between the cell shapes and tensions in cell-cell junctions. This coupling represents an active force which drives the system out of equilibrium and leads to the formation of nematic order interspersed with prominent, long-lived  $+1$ defects. The defects in the nematic ordering are coupled to the shape of the cell tiling, affecting cell areas and coordinations. This intricate interplay between cell shape, size, and coordination provides a possible mechanism by which tissues could spontaneously develop long-range polarity through local mechanical forces without resorting to long-range chemical patterning. 
\end{abstract}

\maketitle

{\textit{Introduction.}}---Epithelial cell monolayers and tissues are prime examples of dense, active, viscoelastic materials \cite{xi2019material,alert2021living,shankar2022topological}. Understanding their behaviour is also of fundamental importance in biology and medicine since epithelia line all organs and cavities in the body, and the majority of cancers develop in epithelial cells~\cite{weinberg2013biology}. The vertex model \cite{honda1983geometrical,farhadifar2007influence,fletcher2014vertex} has played an important role in modelling epithelial mechanics. It can naturally capture the solid-to-fluid transition observed in experiments~\cite{park2015unjamming} by tuning its geometric parameters~\cite{bi2015density}. It is also straightforward to extend it to include active effects, both as self-propulsion~\cite{bi2016motility,barton2017active} and as cell junction activity~\cite{curran2017myosin,krajnc2018fluidization,staddon2019mechanosensitive,krajnc2020solid,sknepnek2021generating,duclut2021nonlinear,yamamoto2022non,duclut2022active}. The vertex model is, therefore, a powerful tool to study active processes in tissues at the cell scale. 

%|||||||||||||||||||||||||||||||||||||||||
%|||||||||||||||figure 1||||||||||||||||||
%|||||||||||||||||||||||||||||||||||||||||
\begin{figure*}[tbh!]
\includegraphics[width=0.99\textwidth]{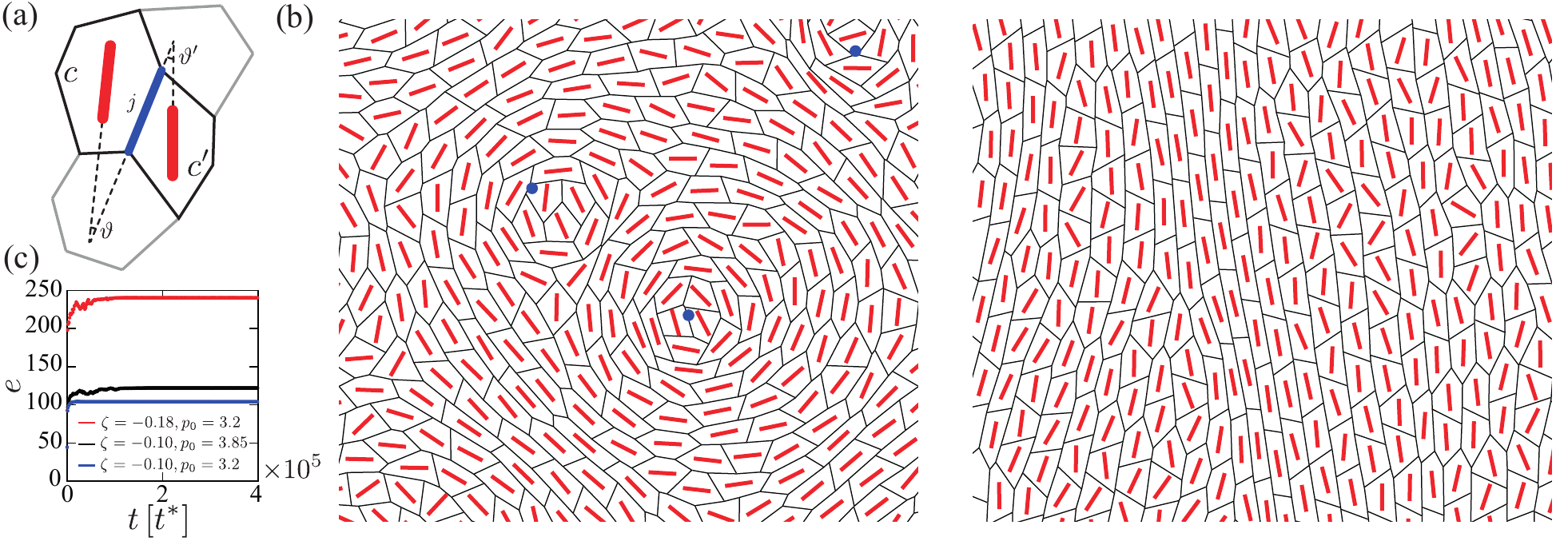}
\caption{\label{fig:model}(a) Schematic representation of the model. Thick red lines indicate the orientations of cell directors. The tension of an edge $j$ (blue line) shared between the two cells depends on its alignment with the cell directors. (b) Snapshots of a simulation of the vertex model with $\zeta=-0.18$ at time $t=10^3$ (left) and $t=10^4$ (right). Cell directors are shown in red, and cores of $+1$ nematic defects are shown as blue disks. In the long time limit, defects can disappear. In panel (b), for clarity, only a quarter of the actual simulation box is shown. (c) Tissue energy as a function of time, including the $\gamma(t) l$ term. } 
\end{figure*}

In many epithelial systems, one also observes nematic-like features at scales that span several cells~\cite{duclos2014perfect,saw2017topological,duclos2017topological,morales2019liquid,eckert2022hexanematic}. These are a readout of elongated cell shapes, and the defects associated with the nematic state have been argued to play important biological roles, e.g.~as sites of cell extrusions~\cite{saw2017topological} or stress-organising centres that drive tissue morphogenesis~\cite{guillamat2022integer}. Theories of active nematics~\cite{simha2002hydrodynamic,marchetti2013hydrodynamics,giomi2015geometry,doostmohammadi2018active} have also been successful in explaining large-scale collective behaviours such as active turbulence both in epithelial cell monolayers~\cite{thampi2016active,mueller2019emergence}, bacterial suspensions~\cite{aranson2022bacterial} and cell membranes~\cite{tan2020topological}.

Cell motion in biological systems is, however, rarely turbulent, and instead one often observes coordinated motion over distances much larger than the typical cell size. Such coordinated movements are key, e.g.~during embryonic development~\cite{wolpert2015principles,rozbicki2015myosin,gros2009cell}. One of the central open questions is how these motions are generated, sustained, and regulated. A closely related question is to what extent such large-scale features require guidance by biochemical patterning, e.g.~via spatio-temporal coordination of morphogens, or whether they can spontaneously emerge as a result of cellular behaviours. It is, therefore, important to understand how cell-level processes coordinate to form tissue- and organ-scale structures.

It has recently been shown that coupling between tension and a global nematic field leads to active T1 transitions that drive tissue shape changes and can elongate cells~\cite{duclut2022active}. In this paper, we explore how nematic order can emerge in a vertex model by introducing coupling between the local cell shape, a proxy for the nematic director, and the tension on cell-cell junctions. 
 We find that this model also leads to prominent $+1$ defects in the nematic order, and we primarily focus on the role of these defects in determining local cell shapes and global tissue tiling.

{\textit{Model details.}}---We begin with the vertex model for planar epithelia~\cite{farhadifar2007influence}. Each cell is represented by a polygon, two cells share a junction, and three or more cells meet at a vertex. The dynamics of the vertices is described by overdamped equations of motion,
\begin{equation}\label{eq:motion}
    \eta \dot{\mathbf{R}}^{(i)} = - \nabla_{\mathbf{R}^{(i)}} E_\text{VM} + \mathbf{F}^{(i)}_\text{act}.
\end{equation}
Here, $\mathbf{R}^{(i)}$ is the position of the $i-$th vertex, the overdot denotes the time derivative, $\nabla_{\mathbf{R}^{(i)}}$ indicates the gradient with respect to $\mathbf{R}^{(i)}$, $E_\text{VM}$ is the elastic energy of the model tissue, $\eta$ is the viscosity, and $\mathbf{F}^{(i)}_\text{act}$ is the active force on the vertex due to coupling between cell shape and junctional tension. The explicit form of $\mathbf{F}^{(i)}_\text{act}$ is discussed below. 

The elastic energy of the vertex model takes the form 
\begin{equation}
    E_{\rm VM}=\sum_{c \in \mathcal{C}} K_A \left(A^{(c)} - A_0\right)^2 + K_P \left(P^{(c)} - P_0\right)^2,
\end{equation}
where the sum is over all cells. $A^{(c)}$ and $P^{(c)}$ are, respectively, the area and perimeter of the cell $c$, whereas $A_0^{(c)}$ and $P_0^{(c)}$ are its target area and perimeter, which we assume to be identical for all cells. $K_A$ and $K_P$ are the area and perimeter elasticity moduli. 

To recast the expression for energy into a dimensionless form, we divide it by $K_AA_0^2$, which gives
\begin{equation}\label{eq:dimensionless}
    e_{\rm VM}=\sum_{c \in \mathcal{C}} \left(a^{(c)} - 1\right)^2 + k_p\left(p^{(c)} - p_0\right)^2,
\end{equation}
where $e_{\rm VM}$ is the dimensionless energy, measured in units of $K_AA_0^2$, and $a^{(c)}=A^{(c)}/A_0$ is the dimensionless area of the cell $c$, measured in units of $A_0$. %$\left(a^*\right)^2$. 
This sets the unit of length as $a^*=\sqrt{A_0}$. Finally, $p^{(c)}=P^{(c)}/\sqrt{A_0}$, $k_P=K_P/(K_AA_0)$, and the target cell-shape index is defined as $p_0=P_0/\sqrt{A_0}$. The parameter $p_0$ plays an important role in determining the mechanical response of the vertex model. Namely, if $p_0$ exceeds a critical value, the model undergoes a solid-to-fluid transition~\cite{bi2014energy}. For regular hexagonal tilings, this occurs at $p_0^c=\sqrt{8\sqrt{3}} \approx 3.72$, while for a random tiling, this occurs at a critical value $p_0^c\approx3.81-3.92$~\cite{merkel2019minimal,wang2020anisotropy,tong2021linear}.

If time is measured in units of $t^*=\eta/\left(K_A A_0\right)$, and force in units of $f^*=K_AA_0^{3/2}$, the equations of motion~\eqref{eq:motion} can be recast into a dimensionless form, 
\begin{equation}\label{eq:dimensionless_motion}
    \dot{\mathbf{r}}^{(i)} = - \nabla_{\mathbf{r}^{(i)}} e_\text{VM} + \mathbf{f}^{(i)}_\text{act}, 
\end{equation}
where $\mathbf{r}^{i}$ is the dimensionless position of the $i-$th vertex and the overdot and $\nabla_{\mathbf{r}^{(i)}}$ now, respectively, indicate the derivative with respect to dimensionless time and the gradient with respect to $\mathbf{r}^{(i)}$.

The dimensionless active force takes the form
\begin{equation}
    \mathbf{f}_\text{act}^{(i)}=-\sum_{j \in \mathcal{J}}\gamma^{(j)}(t)\nabla_{\mathbf{r}^{(i)}} l^{(j)},
    % \mathbf{f}_\text{act}^{(i)}=\nabla_{\mathbf{r}^{(i)}}\sum_{j \in \mathcal{J}}\gamma^{(j)}(t) l^{(j)},
\end{equation}
where the sum is over the set $\mathcal{J}$ of all cell-cell junctions, $\gamma^{(j)}(t)$ is the tension of the $j-$th junction at time $t$, and $l_j$ is its length. The tension in a junction evolves according to 
\begin{equation}
\label{eq:relaxation}
    \dot{\gamma}^{(j)}(t)=-\frac{1}{\tau_\gamma}\left(\gamma^{(j)}-\gamma^{(j)}_0\right),
\end{equation}
where $\tau_\gamma$ sets a characteristic relaxation time scale and $\gamma^{(j)}_0$ is the target tension of the junction. Here, it is selected  to couple the tension in the junction with the elongation of its neighbouring cells, by choosing
\begin{equation}
    {\label{eq:coupling}
    \gamma_0^{(j)} = -\frac{1}{2}\zeta \left[\cos\left(2 \vartheta\right)+\cos\left(2 \vartheta'\right)\right].
    }
\end{equation}
$\vartheta$ and $\vartheta'$ are the angles between junction $j$ and the directors of its neighbouring cells $c$ and $c'$ (Fig.~\ref{fig:model}a), and $\zeta$ is a coupling constant. The sign of $\zeta$ determines whether the active forces act to extend or contract a junction that is aligned with the cell's director. We define a cell's director to point along the eigenvector corresponding to the largest eigenvalue of the cell's gyration tensor, given as
\begin{equation}\label{eq:gyration}
    \mathbf{G}=\frac{1}{n^{(c)}}\sum_{i \in \mathcal{V}_c} \left(\mathbf{r}^{(i)}-\mathbf{r}_0^{(c)}\right)\otimes \left(\mathbf{r}^{(i)}-\mathbf{r}_0^{(c)}\right).
\end{equation}
Here the sum is over the set $\mathcal{V}_c$ of $n^{(c)}$ vertices of the cell $c$, and $\mathbf{r}_0^{(c)}=(1/n^{(c)})\sum_{i\in\mathcal{V}_c}\mathbf{r}^{(i)}$, is the position of the cell's geometric centre. 
Cells for which the gyration tensor has two identical eigenvalues do not contribute to Eq.~\eqref{eq:coupling}.

It is important to note that the choice of the coupling between the cell geometry and junction tensions is key for making the system active. Namely, for the above model,  the active force cannot be written as a gradient of a line tension contribution to the energy, which would lead to the dynamics of the system corresponding to passive energy minimisation. 
Instead, the tensions $\gamma^{(j)}$ are coupled to the instantaneous geometry through Eqs.~\eqref{eq:coupling} and \eqref{eq:relaxation}, but the resulting forces are \emph{only} along the junctions. The movement of the vertices they produce can and does change the tissue shape in such a way that the coupling in Eq.~\eqref{eq:coupling} further increases the energy because angles $\theta$ and $\theta'$ have changed. This is qualitatively different from inserting $\gamma_0$ from Eq.~\eqref{eq:coupling} directly into an energy of the form $\gamma^{(j)}_0 l^{(j)}$, in which case the gradient would lead to additional terms that would rotate junctions, and the model tissue would relax towards a local energy minimum. Effectively, in our model, the gradient does not ``know'' that the angles $\theta$ and $\theta'$ are included in the line tension. As a result, the movement of the vertices does not in fact minimise the energy, rendering the system active. Figure~\ref{fig:model}c shows how the energy of the model tissue changes in time. 

We solve the equation of motion [Eq.~\eqref{eq:dimensionless_motion}] using the first-order Euler scheme with the time step $\delta t=0.01$. In all simulations, we set $k_p=0.02$ and $\tau_{\gamma}=1$. We start simulations with a hexagonal lattice in which each vertex is perturbed by a displacement vector $\delta \mathbf{r}^{(i)}=\delta r^{(i)} (\cos(\alpha^{(i)}),\sin(\alpha^{(i)}))$, where $\delta r^{(i)}$ and $\alpha^{(i)}$ are uniformly distributed random numbers respectively drawn from the intervals $\left[0,0.1\right]$ and $[0,2\pi]$. This is necessary to generate non-zero active forces that otherwise vanish for regular hexagonal tiling. Unless otherwise specified, we used a $32\times32$ lattice of cells placed in a periodic simulation box of fixed size. The longest simulations were run until $t_\text{max}=4 \cdot 10^5$.

Finally, to capture plastic events (i.e.~local cell rearrangements), we implemented T1 transitions~\cite{fletcher2013implementing}. These are neighbour exchanges facilitated by the shrinking of a junction shared by two cells and then creating a new junction between cells that were previously separated. T1 transitions are implemented as follows: If the length of a junction falls below a threshold value $l_0=0.01$ and has decreased since the previous time step, the junction is merged into a four-fold vertex. The four-fold vertex is maintained for $\Delta t = 0.03$. It is then resolved into the new configuration, separating the 
two newly created vertices by $l'=0.001$. Immediately following this resolution, the tension of the new junction is set to 0. Note that our implementation does not allow for a vertex with more than four neighbours, so a junction that is connected to a four-fold vertex cannot undergo a T1 transition.

%%%%%%%%%%%%%%%%%%%%%%%%%%%%%%%%%%%%%%%%%%%%%%%%%%%%%%%%%%%%%%%%%%%%%%%%
%%%%%%%%%%%%%%%%%%%%%%%%%%%%%%%%%%%%%%%%%%%%%%%%%%%%%%%%%%%%%%%%%%%%%%%%
%%%%%%%%%%%%%%%%%%%%%%%%%%%%%%RESULTS%%%%%%%%%%%%%%%%%%%%%%%%%%%%%%%%%%%
%%%%%%%%%%%%%%%%%%%%%%%%%%%%%%%%%%%%%%%%%%%%%%%%%%%%%%%%%%%%%%%%%%%%%%%%
%%%%%%%%%%%%%%%%%%%%%%%%%%%%%%%%%%%%%%%%%%%%%%%%%%%%%%%%%%%%%%%%%%%%%%%%

{\textit{Results.}}---We 
focus on the case $\zeta<0$, i.e.~where junctions are under a higher tension if they are aligned with local cell elongation. Above a threshold value $\zeta_c$, cell shapes are irregular polygons, but all cells have the six-fold coordination of the initial configuration. For $\zeta<\zeta_c$, however, a local alignment of elongated cells starts to develop (Fig.~\ref{fig:model}b). 
The absolute value of the threshold $\zeta_c$ at which local nematic order emerges decreases with increasing $p_0$. 

%|||||||||||||||||||||||||||||||||||||||||
%|||||||||||||||figure 2||||||||||||||||||
%|||||||||||||||||||||||||||||||||||||||||
\begin{figure}[t]
\includegraphics[width=\columnwidth]{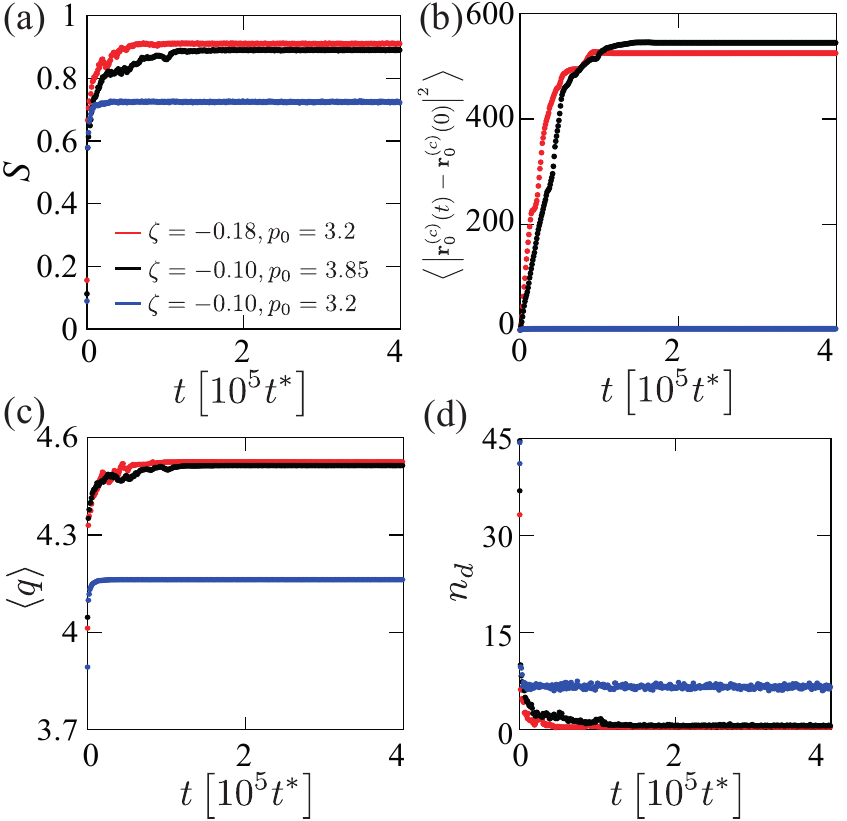}
\caption{\label{fig:nematic} Time-dependence of (a) nematic order parameter $S$, (b) mean square displacement of cell centres, (c) average cell-shape index $q$ of the tissue, (d) number of $+1$ nematic defects.  All panels show data for the same three parameter sets, and the values  obtained are averaged over 10 simulation runs.
}
\end{figure}
 
To quantify the emerging nematic order, we calculated the order parameter
\begin{equation}
    S=\langle\cos\left(2\beta_{c',c}\right)\rangle,
\end{equation}
where $\left<\cdots\right>$ denotes 
an average over all pairs of adjacent cells $c$ and $c'$ and $\beta_{c',c}$ is the angle between the directors of those cells. The time evolution of $S$ is shown in Fig.~\ref{fig:nematic}a. The two parameter sets with high final values of $S$ (red and black lines) correspond to tissues with a significant cell motion, and global nematic order with few or no defects eventually emerges in the majority of simulation runs. However, for lower $|\zeta|$ or $p_0$ (illustrated by $\zeta=-0.1$, $p_0=3.2$, shown in blue in Fig.~\ref{fig:nematic}a) cells do not move significantly and defects in the nematic ordering do not disappear from the tissue.

To illustrate this 
further, we calculated the mean square displacement of cells from their initial positions as a function of time,

\begin{equation}
    {\rm MSD}(t)=\frac{1}{N_\mathcal{C}}\sum_{c\in\mathcal{C}}\left|\mathbf{r}_0^{(c)}(t)-\mathbf{ r}_0^{(c)}(0)\right|^2,
\end{equation}
where $N_\mathcal{C}$ is the total number of cells and $\mathbf{ r}^{(c)}_0(t)$ is the position of the centre [defined below Eq.~\eqref{eq:gyration}] of cell $c$ at time $t$, whereas $\mathbf{r}^{(c)}(0)$ is its initial position at the start of the simulation. The resulting plot is shown in Fig.~\ref{fig:nematic}b. 
While the movement of the tissue is for the most part arrested once the fully nematically ordered state develops, it is important to note that due to active tension coupling this state  is not necessarily the result of energy minimisation.

This nematic-like state also features very high values of the cell-shape index,
\begin{align}\label{eq:shape}
    \langle q\rangle=\frac{p^{(c)}}{\sqrt{a^{(c)}}},
\end{align}
as shown in Fig.~\ref{fig:nematic}c. In all cases, we find average cell-shape indices well above 3.81, which is generally associated with a fluid-like tissue. This remains the case even after cell movement has been arrested.

%|||||||||||||||||||||||||||||||||||||||||
%|||||||||||||||figure 3||||||||||||||||||
%|||||||||||||||||||||||||||||||||||||||||
\begin{figure}[b]
\includegraphics[width=\columnwidth]{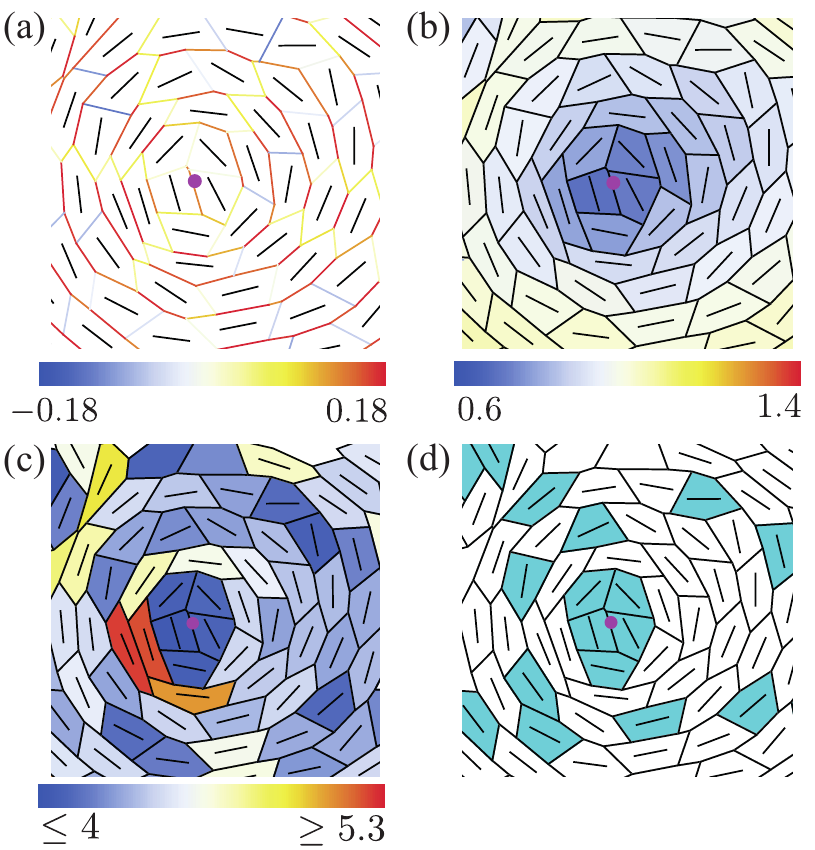}
\caption{\label{fig:zoom}Zoom in on a $+1$ nematic defect from Fig.~\ref{fig:model}b (defect core shown in red). Cells are coloured as follows: (a) with edges coloured according to the tension arising from the shape-tension coupling, (b) according to the cell area, (c) according to the cell-shape index [Eq.~\eqref{eq:shape}], (d) with cells with five neighbours in cyan and others in white. Cell directors are shown in black on all panels.
}
\end{figure}

%%%%%%%%%%%%%%%%%%%%%%%%%%%%%%%%%%%%%%%%%%
%%%%%%%%%%%%%%%NEMATIC ORDER%%%%%%%%%%%%%%
%%%%%%%%%%%%%%%%%%%%%%%%%%%%%%%%%%%%%%%%%%

A striking feature of the cell configurations is the presence of prominent, vortex-like $+1$ defects where cells form concentric rings around the defect core (Fig.~\ref{fig:model}b). These defects either remain for the entire duration of the simulation or vanish over time, for parameter values that lead to the global nematic order. 
Figure~\ref{fig:nematic}d shows the number of $+1$ defects as a function of time. 

While  $+1$ defects are featured most prominently, defects with different topological charges are also present and can therefore be involved in the annihilation of $+1$ defects. It should, however, be stressed that cells in the vertex model are not hard rods and can deform in such a way that the director changes discontinuously, e.g.~going through an isotropic intermediary shape for which the gyration tensor [Eq.~\eqref{eq:gyration}] has two identical eigenvalues and therefore the cell has no director. Therefore, topological charge is not necessarily strictly conserved.

%|||||||||||||||||||||||||||||||||||||||||
%|||||||||||||||figure 4||||||||||||||||||
%|||||||||||||||||||||||||||||||||||||||||
\begin{figure}[tbh!]
\includegraphics[width=\columnwidth]{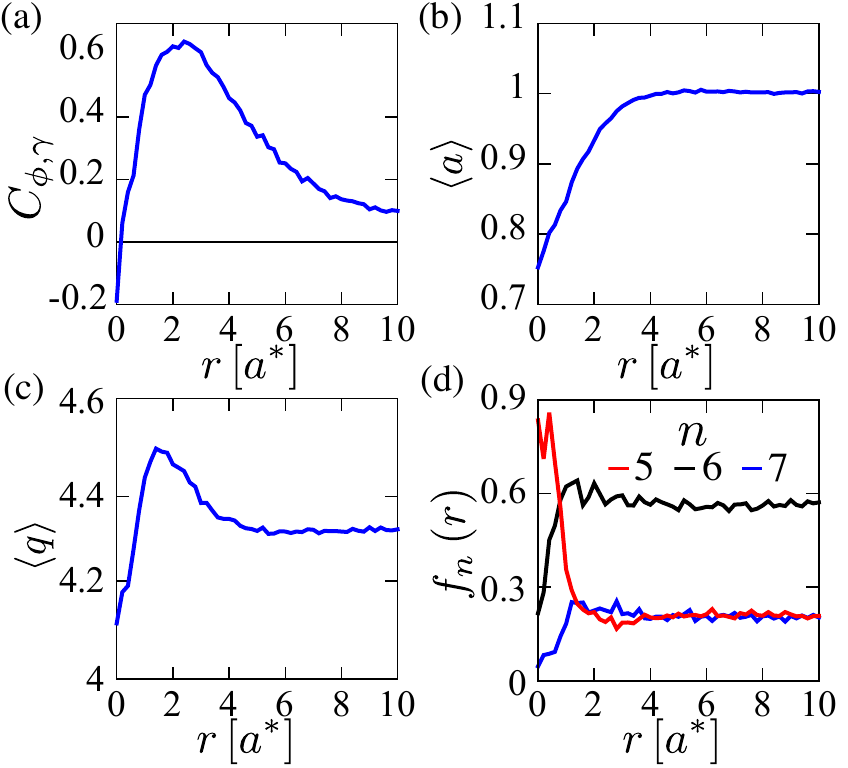}
\caption{\label{fig:defects} Dependence on distance $r$ from the core of a $+1$ defect of (a) angle-tension correlation function, $C_{\phi,\gamma}(r)$ (Appendix \ref{app:elongation}), (b) average cell area, (c) average cell-shape index, (d) distribution of 5-, 6-, and 7- sided cells. All panels are based on simulations starting with a $64\times64$ lattice and a bin size $\Delta r = 0.2$, combining data from 20 simulation runs with  $p_0=3.2$ and $\zeta=-0.18$ at time $t=10^3$.
}
\end{figure}

As noted, the negative values of $\zeta$ favour high tension in junctions that are aligned with the cell director. This leads to the formation of concentric circles of high-tension junctions around vortex-like $+1$ defects (Fig.~\ref{fig:zoom}a) that constrict the defect core. To quantify this, we define the angle $\phi$ between the direction of the junction and the radial vector connecting its centre to the defect core. In Fig.~\ref{fig:defects}a we show how the correlation of the angle $\phi$ and the junction tension changes as a function of distance from the defect core (see Appendix~\ref{app:elongation} for details).  
As a result of the concentric circles of high-tension junctions, cells close to the core of a nematic vortex-like $+1$ defect are compressed compared to the cells further away, and the area elasticity balances the junctional tensions (Figs.~\ref{fig:zoom}b and ~\ref{fig:defects}b). This is a plausible explanation for the long life of $+1$ defects, as the circles of high-tension junctions prevent a $+1$ defect from splitting into a pair of $+1/2$ defects. 

The measured cell-shape index   [Eq.~\eqref{eq:shape}] also depends on the position of $+1$ defects (Figs.~\ref{fig:zoom}c and \ref{fig:defects}c). In particular, we find a decrease in $\langle q\rangle$ in the defect core, followed by a maximum moving outwards from the defect. While cells in the defect core generally have smaller areas, they are also more isotropic, leading to smaller perimeters and, therefore, lower shape indices. In turn, the cells near, but outside, the defect core has a more pronounced elongation compared to those further away, leading to a higher $\langle q\rangle$. Directly measuring how elongation changes with distance from the defect core gives a similar profile to that seen in Fig.~\ref{fig:defects}c (see Appendix~\ref{app:elongation}).

The nematic order couples back to the cell tiling through these $+1$ defects. In particular, we find that cells with five neighbours are very common in the vicinity of the $+1$ defect cores (Fig.~\ref{fig:zoom}d). To quantify this, we computed the distribution function,
\begin{equation}
    f_n(r)=\frac{1}{n_d}\sum_{d=1}^{n_d}\frac{N_n^{(d)}(r)}{\pi \left([r+\Delta r]^2-r^2\right)},
\end{equation}
where the sum is over all $n_d$ $+1$ defects in all 20 simulations used, and $N_n^{(d)}(r)$ is the number of $n$-sided cells at distances between $r$ and $r+\Delta r$ from $+1$ defect $d$. 
 The normalisation was chosen such that the two-dimensional integral of the function would give the number of $n-$sided cells in the model tissue if $f_n(r)$ were based on a single simulation run. Interestingly, both +1 defects~\cite{maroudas2021topological,guillamat2022integer} and five-neighbour cells~\cite{rozman2020collective,rozman2021morphologies,hoffmann2022theory} in the tiling have previously been reported as connected to the formation of budded structures in 3D tissues, so their co-location in our model may be relevant, though typically the experimentally reported +1 defects leading to budding are of the aster type.

%%%%%%%%%%%%%%%%%%%%%%%%%%%%%%%%%%%%%%%%%%%%%%%%%%%%%%%%%%%%%%%%%%%%%%%%
%%%%%%%%%%%%%%%%%%%%%%%%%%%%%%%%%%%%%%%%%%%%%%%%%%%%%%%%%%%%%%%%%%%%%%%%
%%%%%%%%%%%%%%%%%%%%%%%%%%%%%%DISCUSSION%%%%%%%%%%%%%%%%%%%%%%%%%%%%%%%%
%%%%%%%%%%%%%%%%%%%%%%%%%%%%%%%%%%%%%%%%%%%%%%%%%%%%%%%%%%%%%%%%%%%%%%%%
%%%%%%%%%%%%%%%%%%%%%%%%%%%%%%%%%%%%%%%%%%%%%%%%%%%%%%%%%%%%%%%%%%%%%%%%

{\textit{Summary and Discussion.}}---
In this paper, we analysed a vertex model extended to include coupling between the elongation of cells and junctional tensions, leading to an active force on the vertices. For appropriate values of parameters, the model tissue forms nematic ordering of elongated cells which, surprisingly, features prominent vortex-like $+1$ defects. Experimentally,  $+1$ defects are less common in biological systems than $\pm1/2$ defects, but they have been reported both \textit{in vitro}~\cite{endresen2021topological} and \textit{in vivo}~\cite{bonhoeffer1991iso,wolf1998spontaneous}. It has also been recently argued that $+1$ defects play a role in morphogenesis~\cite{maroudas2021topological,guillamat2022integer}.

Moreover, defects in the nematic order are coupled to the tiling of the confluent tissue. 
Cells around the $+1$ defects arrange themselves in such a way that the defects are surrounded by nearly concentric circles of high-tension junctions. This leads to the compression of cells near the defect cores. Moreover, cells that form the defect cores typically have five neighbours. 

An alternative way to introduce active dipolar forces into the vertex model has been proposed in Ref.~\cite{duclut2022active}. 
The key difference to our approach is that tensions in our model are coupled directly to the local elongation of cells, rather than to an external, uniform nematic field. This allows the active tensions to reorient the nematic field to which they are coupled and does not require global patterning to drive the formation of nematic order. Furthermore, recent studies have proposed mechanisms by which extensile stresses can arise in a purely contractile nematic system due to polar fluctuating forces~\cite{vafa2021fluctuations,killeen2022polar} or active  inter-cellular interactions~\cite{zhang2021active} Such mechanisms could provide insights into the behaviour of epithelial monolayers on a substrate.

We also note that the emergence of nematic order due to activity arising from coupling with the local cell elongation has been reported in Ref.~\cite{lin2022tissue}. That study introduces activity via an active stress term proportional to a  $\mathbf{Q}$ tensor, constructed from the  positions of the junctions of each cell. This results in different active forces on vertices. 
It  leads to both tissue fluidisation 
and the emergence of nematic order with $\pm1/2$ defects, for sufficiently high activity.  

Regarding the effects of noise that are inherently present in all biological systems, the movement of cells in our model arises directly from the shape-tension coupling [Eqs.~\eqref{eq:coupling} and \eqref{eq:relaxation}] and does not rely on noise modelled as an Ornstein–Uhlenbeck process used, e.g.~in Refs.~\cite{curran2017myosin,krajnc2020solid,duclut2022active}. A noise term was, therefore, omitted from our model.

Finally, the model studied here assumes a specific form of coupling between cell shape and junctional tension. 
As discussed, this results in an active force that leads 
 to rich collective behaviours. It is, therefore, important to ask how such local coupling could arise in real tissues.  There is evidence that cells can sense their shape~\cite{haupt2018cells}. In, e.g.~the fly, large-scale chemical patterning of cytoskeletal molecules is observed~\cite{bertet2004myosin,bosveld2012mechanical,tetley2016unipolar} that gives global directionality to the tissue, making the model of Duclut, et al.~\cite{duclut2022active} applicable. On the other hand, in systems such as early-stage avian embryos~\cite{rozbicki2015myosin}, there is no such global patterning, yet local anisotropy of cell shapes and actomyosin orientation is apparent, albeit with no clear nematic order. While the molecular details of such coupling are likely to be very intricate, this suggests that it is plausible to consider a scenario in which a cell can locally inform its junctions about its current direction.

%%%%%%%%%%%%%%%%%%%%%%%%%%%%%%%%%%%%%%%%%%%%%%%%%%%%%%%%%%%%%%%%%%%%%%%%
%%%%%%%%%%%%%%%%%%%%%%%%%%%%%%%%%%%%%%%%%%%%%%%%%%%%%%%%%%%%%%%%%%%%%%%%
%%%%%%%%%%%%%%%%%%%%%%%%%%%%%%ACKNOWLEDGMENTS%%%%%%%%%%%%%%%%%%%%%%%%%%%
%%%%%%%%%%%%%%%%%%%%%%%%%%%%%%%%%%%%%%%%%%%%%%%%%%%%%%%%%%%%%%%%%%%%%%%%
%%%%%%%%%%%%%%%%%%%%%%%%%%%%%%%%%%%%%%%%%%%%%%%%%%%%%%%%%%%%%%%%%%%%%%%%
\begin{acknowledgments}
We wish to thank Mehrana R.~Nejad and Guanming Zhang for insightful discussions, and Matej Krajnc for providing the initial version of the vertex model code. J.R.~and J.M.Y. acknowledge support from the UK EPSRC  (Award EP/W023849/1). R.S. acknowledges support from the UK EPSRC (Award EP/W023946/1).
\end{acknowledgments}

%%%%%%%%%%%%%%%%%%%%%%%%%%%%%%%%%%%%%%%%%%%%%%%%
%%%%%%%%%%%%%%%%%%%%%%%%%%%%%%%%%%%%%%%%%%%%%%%%
%%%%%%%%%%%%%%%%%%APPENDIX%%%%%%%%%%%%%%%%%%%%%%
%%%%%%%%%%%%%%%%%%%%%%%%%%%%%%%%%%%%%%%%%%%%%%%%
%%%%%%%%%%%%%%%%%%%%%%%%%%%%%%%%%%%%%%%%%%%%%%%%
\appendix

\section{Defect detection}\label{app:detection}
In order to locate defects in the nematic order, we calculated the change in the orientation of the nematic order around a vertex and then merged nearby vertices with the same non-zero charge to determine defect positions. Specifically, the defect-finding algorithm is as follows: For each vertex, we find its third-nearest-neighbour cells. We order those cells counter-clockwise based on the angle between the vector pointing from the vertex to the cell  and the horizontal axis of the simulation box. We then loop over the cells and calculate the winding number around the vertex~\cite{killeen2022polar},
\begin{equation}
    \Delta \alpha^{(c)} =\sum_{c}^{m_3^{(i)}}\delta \alpha_{c,c+1},
\end{equation}
where the sum is over all  $m_3^{(i)}$  cells $c$, in counterclockwise order, that are third-nearest-neighbours of vertex~$i$. $\delta \alpha_{c,c+1}$ is the angle between the directors of cells $c$ and $c+1$, where we take into account that these vectors are headless, which limits the angles to $\left[-\pi/2,\pi/2\right]$. The charge at the vertex is then $\Delta \alpha^{(i)}/(2\pi)$. For each vertex, we then find all other vertices within three cells' distance: If multiple have the same charge, the defect core is defined as the average position of those vertices (the final result may slightly depend on the order in which the vertices are considered for this step). We found that this approach is generally more accurate for finding +1 defects compared to the grid-based approaches implemented in, e.g.~Refs.~\cite{lin2022tissue,killeen2022polar}.

This algorithm does not distinguish between  vortex- and aster-shaped defects. Both are therefore included in calculations for Figs.~\ref{fig:defects} and \ref{fig:elongation}. Aster-shaped defects are, however, not common, so the statistics pertain chiefly to vortex-like $+1$ defects.

\section{Tension-angle correlation and cell elongation}\label{app:elongation}

\begin{figure}[t]
\includegraphics[width=0.95\columnwidth]{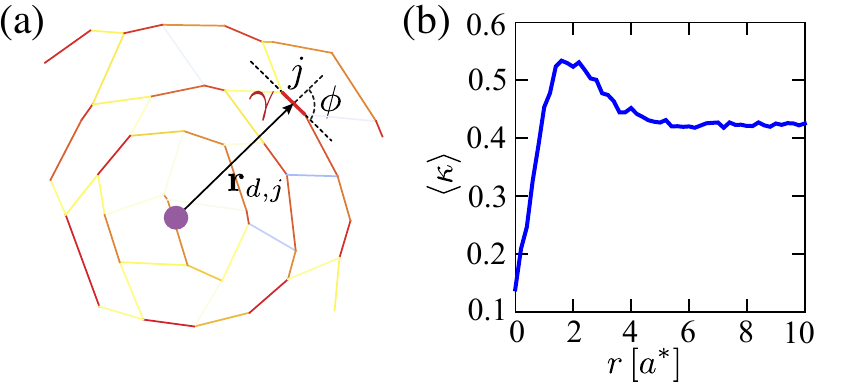}
\caption{\label{fig:elongation} (a) Schematic representation of the ingredients used to construct the correlation function in Fig. \ref{fig:defects}a. Junctions are coloured as in Fig.~\ref{fig:zoom}a. (b) Average cell elongation as a function of distance from $+1$ nematic defects, combining data from 20 simulations with $p_0=3.2$ and $\zeta=-0.18$ at time $t=10^3$, for a $64\times64$ lattice.
}
\end{figure}

We compute the correlation function shown in Fig.~\ref{fig:defects}a as follows: For each pair of a +1 defect $d$ and a junction $j$, we first create a vector $\mathbf{r}_{d,j}$ going from the defect to the centre of the junction (Fig.~\ref{fig:elongation}a). We then determine the angle $\phi$ between $\mathbf{r}_{d,j}$ and the junction. As junctions are effectively headless vectors, these angles are limited to $\left[0,\pi/2\right]$. 
The correlation $C_{\phi,\gamma}(r)$ between $\phi$ and the junctional tension $\gamma$ is then calculated using data from all junctions for which $|\mathbf{r}_{d,j}|$ is between $r$ and $r+\Delta r$ (where $\Delta r = 0.2$) for all defect $d$ in 20 simulations.

%|||||||||||||||||||||||||||||||||||||||||
%|||||||||||||||figure 5||||||||||||||||||
%|||||||||||||||||||||||||||||||||||||||||

We calculate the cell elongation as
\begin{equation}
    \kappa=\frac{g_1-g_2}{g_1+g_2},
\end{equation}
where $g_1$ and $g_2$ are, respectively, the larger and smaller eigenvalue of the cell gyration tensor given in Eq.~\eqref{eq:gyration}. Figure ~\ref{fig:elongation}b shows how  $\left<\kappa\right>$ varies as a function of the distance from a $+1$ nematic defect. 

\providecommand{\noopsort}[1]{}\providecommand{\singleletter}[1]{#1}%
%

%\bibliography{nematic_vertex}

\end{document}